# Manipulating the Topology of Nanoscale Skyrmion Bubbles by Spatially Geometric Confinement

Zhipeng Hou[†,‡,*], Qiang Zhang[‡,*], Guizhou Xu[§,*], Senfu Zhang[‡], Chen Gong[‡], Bei Ding[†], Hang Li[†], Feng Xu[§], Yuan Yao[†], Enke Liu[†], Guangheng Wu[†], Xi-xiang Zhang[‡,#], and Wenhong Wang[†,#]

[†]Beijing National Laboratory for Condensed Matter Physics, Institute of Physics, Chinese Academy of Sciences, Beijing 100190, China

[‡]King Abdullah University of Science and Technology (KAUST), Physical Science and Engineering, Thuwal 23955-6900, Saudi Arabia

[§]School of Materials Science and Engineering, Nanjing University of Science and Technology, Nanjing 210094, China

*These authors contributed equally to this work.

[#]To whom correspondence should be addressed to W. H. Wang (email: wenhong.wang@iphy.ac.cn ) or X. X. Zhang (email: xixiang.zhang@kaust.edu.sa)

**Abstract:** The discovery of magnetic skyrmion bubbles in centrosymmetric magnets has been receiving increasing interest from the research community, due to the fascinating physics of topological spin textures and its possible applications to spintronics. However, key challenges remain, such as how to manipulate the nucleation of skyrmion bubbles to exclude the trivial bubbles or metastable skyrmion bubbles that usually coexist with skyrmion bubbles in the centrosymmetric magnets. Here, we report having successfully performed this task by applying spatially geometric confinement to a centrosymmetric frustrated $Fe_3Sn_2$ magnet. We demonstrate that the spatially geometric confinement can indeed stabilize the skyrmion bubbles, by effectively suppressing the formation of trivial bubbles and metastable skyrmion bubbles. We also show that the critical magnetic field for the nucleation of the skyrmion bubbles in the confined $Fe_3Sn_2$ nanostripes is drastically less, by an order of magnitude, than that what is required in the thin plate without geometrical confinement. By analyzing how the width and thickness of the nanostripes affect the spin textures of skyrmion bubbles, we infer that the topological transition of skyrmion bubbles is closely related to the dipole-dipole interaction, which we find is consistent with theoretical simulations. The results presented here represent an important step forward in manipulating the topological spin textures of skyrmion bubbles, making us closer to achieving the fabrication of skyrmion-based racetrack memory devices.

Magnetic skyrmions are topologically protected vortex-like objects that were first discovered in the chiral, non-centrosymmetric magnets[1-6] where they are stabilized by the Dzyaloshinskii-Moriya interaction (DMI). Unlike the conventional, "rigid" magnetic domain walls, they are flexible in shape-deformation to avoid pinning centres,[7,8] and therefore only an ultra-low current density is needed to drive them to move.[7-12] This fascinating topological property, combined with their nanoscale size and stable particle-like features make skyrmions promising candidate for carrying magnetic information in further high-density and low-power consumption spintronic devices based on the racetrack memory concept.[8,13,14-24]

Recent studies showed that some non-chiral centrosymmetric magnets could host skyrmion bubbles (SKBs),[25-32] stabilized by the interplay of the external magnetic field, ferromagnetic exchange interaction, uniaxial magnetic anisotropy, and dipole-dipole interaction (DDI). SKBs are topologically equivalent to magnetic skyrmions and exhibit similar topological properties, such as the topological Hall effect,[27] skyrmion Hall effect,[33] and ultra-low driving current density for current-induced motion.[25] More importantly, SKBs show a significantly high thermal stability over a wide temperature range crossing room temperature,[27,28,34] showing a high potential of SKBs for the construction of memory devices. Contrary to DMI-stabilized skyrmions with a fixed helicity, SKBs in centrosymmetric magnets possess two degrees of freedom, *i.e.*, vorticity and helicity,[35] which makes them usually coexist with the topologically trivial bubbles (topological number is equal to 0),[28-31,37] or exhibit multiple topologies such as biskyrmions,[25,27,30,36] and various metastable SKBs[28,30,36] (for example, pendulum-shaped SKBs[29] and bifurcation-shaped SKBs[28,30,36]). When external stimuli, such as magnetic field $H$ or spin-polarized current, are applied, the spin structures of the trivial and metastable SKBs may vary with the motion of Bloch lines (BLs), making them unsuitable for the application in magnetic racetrack memory devices. Therefore, in order to be suitable for such applications, it is essential to remove trivial bubbles and metastable SKBs.[21]

Recent theoretical simulations, based on the nanostructured frustrated magnet, showed that the periodically modulated spin textures at geometrical boundaries had a significant influence on the magnetization dynamics of SKBs,[35] offering a new path that designing proper geometric confinement to manipulate the topology of SKBs. Moreover, when using a nanostructured magnet, the strong geometric confinement can efficiently suppress the skyrmion Hall motion.[38] Due to the importance of the spatially geometric confinements for both fundamental physics and practical applications of SKBs, it is of great significance to experimentally clarify the influence of geometric confinement on the nucleation of SKBs in nanostructures.

$Fe_3Sn_2$ is a centrosymmetric frustrated ferromagnetic compound which has a layered rhombohedral structure with an alternate stacking of Sn layers and Fe-Sn bilayers along the $c$-axis.[28,39,40] The *in situ* Lorentz transmission electron microscopy (LTEM) observations on a mechanically polished $Fe_3Sn_2$ thin plate[28] revealed that the SKBs coexist with the trivial bubbles and exhibit multiple topologies. Very recently, we further demonstrated that the SKBs in the $Fe_3Sn_2$ nanostripes possess an extremely high temperature stability,[34] which is of great significance for the practical applications of SKBs. Here, we provide a comprehensive study of the effects of spatial confinement on the nucleation of SKBs in nanostripes fabricated from a single crystalline centrosymmetric frustrated magnet $Fe_3Sn_2$. We successfully manipulated the evolution path of SKBs and completely excluded the trivial bubbles or metastable SKBs that coexist with the SKBs in $Fe_3Sn_2$. Importantly, we found that the critical magnetic field for the formation of SKBs decreased drastically from 860 mT in the large thin plate without geometric confinement to merely 70 mT in a nanostripe with a width of 600 nm and thickness of 200 nm. These findings not only offer us a fundamental insight into the physics of effect of spatial confinement on the nucleation of skyrmion bubbles, but also take a significant step towards their application in spintronic devices.

**RESULTS and DISCUSSION**

We first simulated numerically the spin structures in the samples with spatially confined geometries, by varying the width $w$ (3 μm - 300 nm) and thickness $t$ (300 nm - 100 nm) of a $Fe_3Sn_2$ single crystal, based on the experimentally achievable parameters. The details of the simulations can be found in the Methods section. In Figure 1a and b, we present the simulated spin textures for a trivial bubble and for SKB, respectively. In Figure 1a, one can notice that the trivial bubble is composed of a pair of open Bloch lines, resulting in a zero topological number (Figure 1a). In contrast, as seen in Figure 1b, the SKB possesses a nonzero topological number and exhibits a similar spin texture with the skyrmions in the chiral magnets (Figure 1b). Based on the spin textures of the two classes of spin configurations, we simulated their corresponding LTEM images, as described in previous work.[41] When imaged by LTEM, the trivial bubble showed non-convergent arcs (Figure 1c), whereas the SKB appeared in a ring-like pattern (Figure 1d). We found these simulated images to be in accordance with the experimental observations.[28]

The numerical simulations predicted that the width confinement would exhibit a striking effect on the topology of SKBs (Figure S1 and S2), with the thickness confinement showing much less influence (Figure S3). Figure 1e shows the calculated spin structures as a function of the external magnetic field $H$ and width $w$ of the nanostripes. We found that, when decreasing the width of nanostripes, both the trivial bubbles and the metastable SKBs gradually disappeared, whereas the topologically stable SKBs persisted. In particular, when the width fell below 800 nm (including 800 nm), only the SKBs could survive at the thermodynamic equilibrium. Based on the simulations, we hence expected to experimentally exclude the trivial bubbles and metastable SKBs by geometric confinement.

To confirm the theoretical predictions experimentally, we first fabricated a nanostripe with a width $w \sim 600$ nm and thickness $t \sim 250$ nm from a high-quality $Fe_3Sn_2$ single crystal, using a focused ion beam (FIB) (Figure 2a). The left panel of Figure 2b shows a typical scanning transmission electron microscopy (STEM) image of the nanostripe. One can notice that the sample was coated with amorphous carbon (black region) and platinum (grey region) to protect the $Fe_3Sn_2$ layer during the fabrication of the sample and to reduce the interfacial Fresnel fringes in the LTEM image.[15] The selected-area electron diffraction (SAED) pattern taken from the nanostripe displayed a characteristic, six-fold symmetry (right panel of Figure 2b), suggesting the normal direction of the nanostripe is along the [001] axis.

Hereafter, we performed the LTEM observations at 300 K under different external magnetic fields applied along the out-of-plane direction of the nanostripe. Figure 2c-f show the corresponding under-focused LTEM images (Their corresponding over-focused LTEM images are shown in Figure S4). In a zero magnetic field, the sample exhibited a stripe domain structure with an average period $\lambda \approx 380$ nm along the long edge of the nanostripe. The average period was found slightly larger than that ($\lambda \approx 320$ nm) observed in the wide $Fe_3Sn_2$ thin plate.[28] As the external magnetic field increased, the stripe domains underwent a series of dynamical transformations before finally settling into the topologically stable SKBs. Once the SKBs were formed, due to the repulsions between the edge spins and the SKBs in both edges, they tended to move towards the middle of the nanostripe, resulting in a densely assembled single chain along the long axis (Figure 2f). This sharply contrasted with the situation observed in the wide $Fe_3Sn_2$ thin plates in which the nucleation of SKBs take place in an isolated and random manner over the whole sample.[28] This observation led us to think that the width confinement strongly favored, not only the creation of SKBs but also the formation of a single chain of SKBs by self-assembly. Furthermore, we found that the width confinement could also significantly influence the evolution process of SKBs and make it clearly different from that in the wider thin plate.[28] The domains enclosed by the white boxes in Figure 2c-f showed a typical transformation process from a typical stripe domain to a SKB. When increasing the magnetic field, the stripe domain (the one enclosed in the square) gradually shrank into a half-skyrmion-like object first (Figure 2c-e), rather than breaking into numerous trivial bubbles, as observed in the wider plates.[28] By increasing the magnetic field further, the half-skyrmion-like object transformed itself directly into an SKB without undergoing a series of complex motions of the Bloch lines induced by the field (Figure 1f). More interestingly, no trivial bubbles or metastable SKBs were observed over the whole process, indicating that the width confinement completely suppressed their formation.

In addition, we found that the threshold field for the nucleation of SKBs decreased drastically, from a very high value of 860 mT in wide thin plate to 130 mT in a $Fe_3Sn_2$ nanostripe with geometric confinement. More intriguingly, we found that, once the single chain of SKBs with uniform topology was formed in the nanostripe with width confinement, it could persist at a much lower magnetic field, and even at a zero magnetic field (Figure 2g-j). The stabilization mechanism of the zero-field SKBs may be attributed to the pinning effect of the geometric edges and the topological protection.[23]

To gain a comprehensive understanding of the effect of the width confinement on the formation of SKBs, we investigated the domain evolution in nanostripes with various widths, ranging from 4 μm (approximately thirteen times of the diameter *D* of a single SKB) to 200 nm (smaller than *D*) (Figure S5). Detailed LTEM images showing the evolution of the magnetic domains under different fields are shown in Figure S5. Shown in Figure 3a-e are the representative LTEM images obtained on nanostripes with typical widths under their corresponding lower-bound threshold magnetic fields. The summary of the experimental results leads to a similar phase diagram (Figure 3f) to the simulated one, composed of four states: mixed SKBs (coexistence of trivial bubbles, metastable SKBs, and SKBs), SKBs, SSC (single SKB chain) and half-skyrmion-like states (HSKs). We would like to emphasize some interesting features in the LTEM images. The formation of the zigzag chain in the *w* ~ 1.8 μm nanostripe (Figure 3c) suggests that the SKBs intend to form a hexagonal lattice if they are densely packed.[42] By further decreasing the width, a straight, single chain of SKBs gradually formed. It is found that the single chain of the SKBs is stable in the nanosripes with width between 800 nm and 350 nm. We also noticed that, a transverse elliptical distortion of SKBs happened in the nanostripes with width range of 500 ~ 350 nm (Figure 3d and Figure S6), which suggests that the radial symmetry of SKBs can be modified by the width confinement without a loss of their topological nature. This was also observed in the nanostructured chiral FeGe magnet.[16] When the width decreased and became smaller than the diameter of a SKB (approximately 350 nm), the nanostripe was not wide enough to accommodate a full SKB. To reduce the total energy, the SKBs would transform into a half-skyrmion-like objects (Figure 3e and Figure S6). As expected, based on the numerical results, we also found that the threshold magnetic field decreased monotonously with decreasing the width (Figure 1e and Figure 3f). These features confirmed that the width confinement could strongly affect the formation of SKBs. To illustrate this effect more vividly, we draw a schematic diagram to summarize the effect of width confinement, as shown in Figure 3g. It is clearly demonstrated that the width confinement provides us a much more convenient evolutionary path that the stripe domains tactfully jump over the trivial bubbles or polymorphous metastable SKBs and directly assemble a single chain of SKBs under a much lower magnetic field. All these features are extremely advantageous for the application of SKBs in racetrack memory devices.

We investigated further the effects of the thickness confinement on the formation of SKBs in the $Fe_3Sn_2$ nanostripes with a fixed width $w \sim 600$ nm. As we discussed above (Figure 3f), a single chain of SKBs forms in naonostripes with a width range of 800 ~ 350 nm and with a thickness of 250 nm. Shown in Figure 4a and b, are the TEM image and the corresponding schematic of a nanostripe with three segments of different thicknesses ($t \sim 300$, 270, and 200 nm) fabricated using FIB. Figure 4c shows the LTEM images of the three-segment sample under an external magnetic field of 130 mT. We clearly see that the overall morphology or the shape of the SKBs is insensitive to the variation of sample thickness, but that the diameter of the SKBs decreases monotonically with decreasing the thickness (Figure 4d). This phenomenon is in a good agreement to that observed in the $Fe_{0.5}Co_{0.5}Si$[43] and MnNiGa thin plates[44] and can be ascribed the decrease of the threshold (or critical) magnetic field for the formation of SKBs in the thinner samples. This correlation relies on the fact that the diameter of the SKBs decreases with increasing external magnetic field. We indeed observed that, by decreasing the magnetic field, the diameter of the SKBs monotonically increased, as shown in Figure 2g-j. To experimentally confirm this, we studied the skyrmion formation in this three-segment sample by varying the magnetic field. Shown in Figure 4e is the lower-bound threshold field for the formation of SKBs as a function of thickness extracted from our observations, which showed a monotonic decrease of threshold field with decreasing thickness. It should be noted that, in the thinnest sample, $t \sim 200$ nm, the threshold field was only 70 mT. This low magnetic field value was of the same order of magnitude as that in the magnetic mutilayers[18,19,21,45] and the chiral magnet Co-Zn-Mn,[46] but much lower than that in other materials that host room-temperature stabilized SKBs or antiskyrmions (such as $Ni_2MnGa$,[26] $MnNiGa$,[27] and $Mn_{1.4}Pt_{0.9}Pd_{0.1}Sn$[47]). Interestingly, when the thickness was decreased to about 150 nm, with increasing magnetic field, the stripe domains would vanish and directly transform to the magnetic saturation without undergoing the phase of SKBs (Figure S7).

Here, we discuss the possible physical mechanisms of the effect of geometric confinement on the nucleation of SKBs in the $Fe_3Sn_2$ nanostripes. It has been established that half-skyrmion-like states play a crucial role in the creation of skyrmions in the nanostructured chiral magnets, in which the strong DMI results in edge spin states at the boundaries of the sample, and forces the spins to propagate along the edges and warp into helicity-fixed half-skyrmion-like states.[15,16] In this work, we showed that the nucleation of the topologically stable SKBs was also closely related to the presence of the half-skyrmion-like states. However, unlike in chiral magnets, the corresponding formation of these spin configurations in centrosymmetric magnets should originate from the dipole-dipole interactions. The domain evolution processes that we observed experimentally could be well reproduced by micromagnetic simulations, when considering the interplay of the external magnetic field, ferromagnetic exchange interaction, uniaxial magnetic anisotropy, and dipole-dipole interaction. Figure 5a-d represent the simulated magnetization process of the domain, in a nanostripe with a width $w \sim 600$ nm and thickness $t \sim 250$ nm, and illustrate the role played by the dipole-dipole interaction. In a zero magnetic field, the spin helixes organize themselves in a chain-like fashion, along the nanostripe, between the top and bottom edges. At the boundaries of the nanostripe, the spin helixes with opposite directions of magnetization attract each other, whereas those with the similar directions repel each other, as evidenced by the slightly different distances marked by the boxes in Figure 5a. This behavior is a natural consequence of the dipole-dipole interactions. As the external magnetic field increases, the strength of the dipole-dipole interaction increases significantly (Figure 5e) and dipole-dipole interaction acts as an unbalanced torque on the edge spins to make their in-plane components propagate along the edges and connect with neighboring two helixes to minimize the stray field energy. This characteristic makes the nucleation of the half-skyrmion-like state in the width-confined nanostripe a more favorable option than the trivial bubbles in energy. By further increasing the magnetic field, since edge spins can be aligned by magnetic field more easily than the interior spins, the spins at the edges gradually got lifted away from the boundaries, resulting in the formation of half-skyrmion-like states that further stabilize the SKBs. A further analysis of the magnetization process of a nanostripe for $w \sim 600$ nm, $t \sim 250$ nm was performed, using the transport-of intensity equation (TIE) (Figure S8). We found the results to be consistent with the micromagnetic simulations, confirming the dipole-dipole interaction as the dominant effect on the nucleation of a half-skyrmion-like-state and thus the formation of SKBs in the geometrically confined $Fe_3Sn_2$ nanostripes.

On the other hand, a decrease of the crucial field for the nucleation of SKBs in the spatially confined nanostripe is closely related to the formation of a half-skyrmion-like-state. As mentioned above, the SKBs in the wider, thin plates are transformed from the trivial bubbles. In this case, the trivial bubbles need to overcome a large potential energy barrier to switch their topological number from zero to one; therefore, an extremely high external magnetic field of 860 mT is required to induce the series of complex motions of Bloch lines. However, in the case of the nanostripe, the width confinement effect promotes the formation of the half-skyrmion-like-state. By analyzing the spin texture, we found the half-skyrmion-like-state to be topologically homeomorphous with the SKB, and that it could be viewed as its elongated form. These novel features make the corresponding energy barrier between the half-skyrmion-like state and SKB much lower than that resulting from the transformation of a trivial bubble into SKB. Therefore, we found that the corresponding critical magnetic field for the nucleation of SKBs decreased drastically, from 860 mT to merely 130 mT, for nanosripe of $w \sim 600$ nm, and $t \sim 250$ nm. By further applying the thickness confinement, the critical field could even be further decreased to 70 mT. This decrease can be explained as follows: a reduction of the sample's thickness increases the in-plane component of spins, which reduces the demagnetization field, significantly suppressing the stability of the conical arrangement of spins along the $z$-axis, and decreasing the Zeeman energy required for the formation of SKBs (Figure S3).

**CONCLUSIONS**

In summary, by employing a combination of micromagnetic simulations and LTEM experimental observations, we have systematically evaluated the effects of geometric confinement on the nucleation process of SKBs, in the centrosymmetric frustrated $Fe_3Sn_2$ magnet. Our results show that the spatial confinement can exclude the trivial bubbles or metastable SKBs while stabilize topologically stable SKBs, in a relatively low critical magnetic field, at room temperature. This represents a significant step towards the application of SKBs for the fabrication of racetrack memory devices. Controlling the topology of magnetic bubbles in geometrically confined nanostructures allows us to design proper spatial confinement to integrate multiple spin textures, opening new avenues for designing new types of spintronic devices, based on their multifarious topologies. Theoretical simulations have revealed that the dipole-dipole interaction also plays a crucial role in manipulating the topology of SKBs, offering valuable insights into the physics and fundamental mechanisms underlying spatial confinement and its effects on the magnetization dynamics of the non-chiral magnets.

**METHODS/EXPERIMENTAL**

**Sample preparation.** High-quality $Fe_3Sn_2$ single crystals were synthesized following a high-temperature flux method[27]. Structural and transport measurements were performed in order to examine the quality of the single crystals. The nanostripes were fabricated from a $Fe_3Sn_2$ single crystal, using a focus ion beam (FIB) equipped with a scanning electron microscope (SEM) dual beam system (Helios NanoLab 400s). The fabrication process included the following steps: (i) a thin plate with a thickness of e.g. 600nm was caved on the (001) surface of a $Fe_3Sn_2$ single crystal, using the FIB milling method; (ii) layers of C and Pt were deposited on the surface of the thin plate; using a Gas Injection System (GIS) to protect the edge of the nanostripe for the nano-manipulation process; (iii) a cuboid was cut from the thin plate by FIB milling; (iv) the cuboid was transferred to a TEM Cu chip, in parallel with the horizontal plane; . (v) the TEM Cu chip was rotated by a 90 degree with the Cu chip perpendicular to the horizontal plane and the cuboid was thinned along the horizontal plane.

**LTEM measurements and TIE analysis.** The magnetic domain structure was detected using a Titan G2 60–300 (FEI), in the Lorentz TEM mode, equipped with a spherical aberration corrector for an imaging system, at an acceleration voltage of 300 kV. In order to determine the spin helicity of the skyrmions, three sets of images with either lower, higher, or exact values of focal lengths were recorded, using a charge-coupled device (CCD) camera the high-resolution in-plane magnetization distribution mapping was then obtained by a QPt software, based on the transport of the intensity equation (TIE). The objective lens was turned off when the sample holder was inserted, and the perpendicular magnetic field was applied to the sample by increasing the objective lens, gradually, in very small increments.

**Micromagnetic simulations.** Micromagnetic simulations were carried out using a three-dimensional object- oriented micromagnetic framework (OOMMF) code, based on the LLG function.[49] The width of the cuboid model varied from 300 nm to 3 μm, whereas the length was kept at 2.4 μm. The material parameters were chosen according to the experimental values of $Fe_3Sn_2$, with a saturation magnetization $M_s$ = 5.66×10$^5$ A/m at room temperature, and a uniaxial magnetocrystalline anisotropy constant $K_u$ = 0.3×10$^5$ J/m$^3$. The value of the exchange constant $A$ we used was 1.4×10$^{-11}$ J/m. The rectangle mesh size was 5 nm × 5 nm × 5 nm, within the exchange length of $l_{ex}$ = 8.3 nm ($l_{ex} = \sqrt{\frac{2A}{\mu_0 M_S^2}}$). To simulate the evolution of stripe domains, we set an initial magnetic state of stripe domains in the $c$-axis ($z$-direction), with a periodicity of 400 nm, and subsequently relaxed it to the equilibrium state by integrating the LLG equation. A damping constant $\alpha$ = 1 was applied to ensure a quick relaxation to the equilibrium state. The evolution of the magnetic field from the domains was achieved by increasing the field upon the previously converged magnetization structure.

## Acknowledgements

This work was supported by the National Key R&D Program of China (Grant Nos. 2017YFA0303202), National Natural Science Foundation of China (Grant Nos. 11604148, 1561145003, 11574374), King Abdullah University of Science and Technology (KAUST) Office of Sponsored Research (OSR) under Award No: CRF-2015-2549-CRG4 and No 2016-CRG5-2977, and the Key Research Program of the Chinese Academy of Sciences，KJZD-SW-M01.

**Supporting Information Available:** <Supporting Information describing the simulated magnetization dynamics in the spatially confined nanostripes and experimentally observed magnetic domain evolution processes in the nanostripes with various widths.> This material is available free of charge *via* the Internet at http://pubs.acs.org.

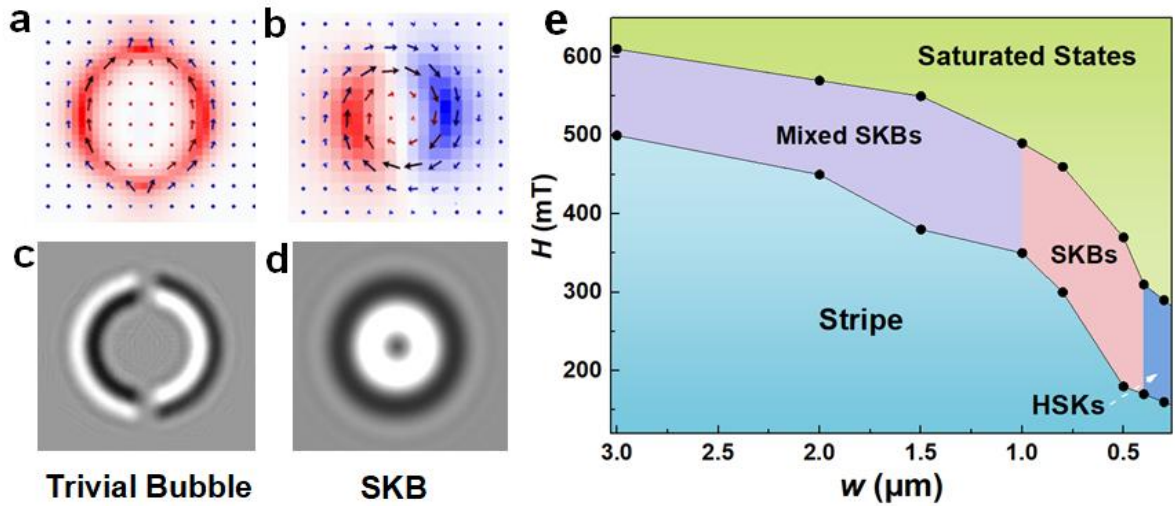

**Figure 1. Calculated spin structure and phase diagram.** Numerical results of the spin structures for **(a)** a trivial bubble and **(b)** a skyrmion bubble (SKB). The simulated LTEM patterns corresponding to **(a)** and **(b)** are shown in **(c)** and **(d)**, respectively. **(e)** The phase diagram of the spin structure as a function of external magnetic field $H$ and width $w$ extracted from the simulation. Depending on the width of the nanostripes, the state of the stripe-domains at the zero magnetic field can transform into four different magnetic states under the appropriate external fields: mixed SKBs (coexistence of trivial bubbles, metastable SKBs, and SKBs), SKBs, and half-skyrmion-like state (HSK).

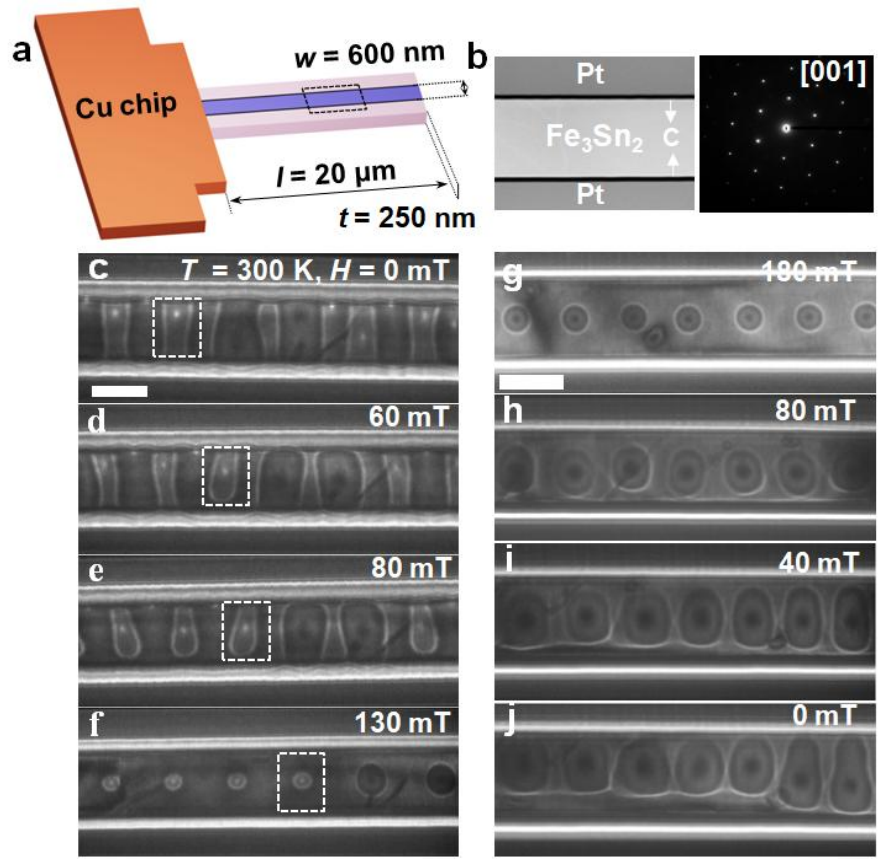

**Figure 2. Evolution of spin structures as a function of the magnetic field for a typical Fe$_3$Sn$_2$ nanostripe.** (a) Schematic view of the Fe$_3$Sn$_2$ nanostripe with a length $l \sim 20$ μm, width $w \sim 600$ nm, and thickness $t \sim 250$ nm; (b) Typical scanning electron microscopy (STEM) image of the 600 nm wide nanostripe (left panel) and its corresponding selected-area electron diffraction pattern (SAED). The six-fold symmetry suggests the beam is along the [001] axis. (c-f) Under-focused LTEM images under different out-of-plane magnetic fields at 300K. The domain enclosed by white boxes present a one-to-one correspondence in the stripe-bubble transformation. (g-j) Variation of the SKBs' morphology with decrease of magnetic field. Notably, the images in **g-j** are taken from the same nanostripe with **c-f** but at different regions. The scale bar is 500 nm.

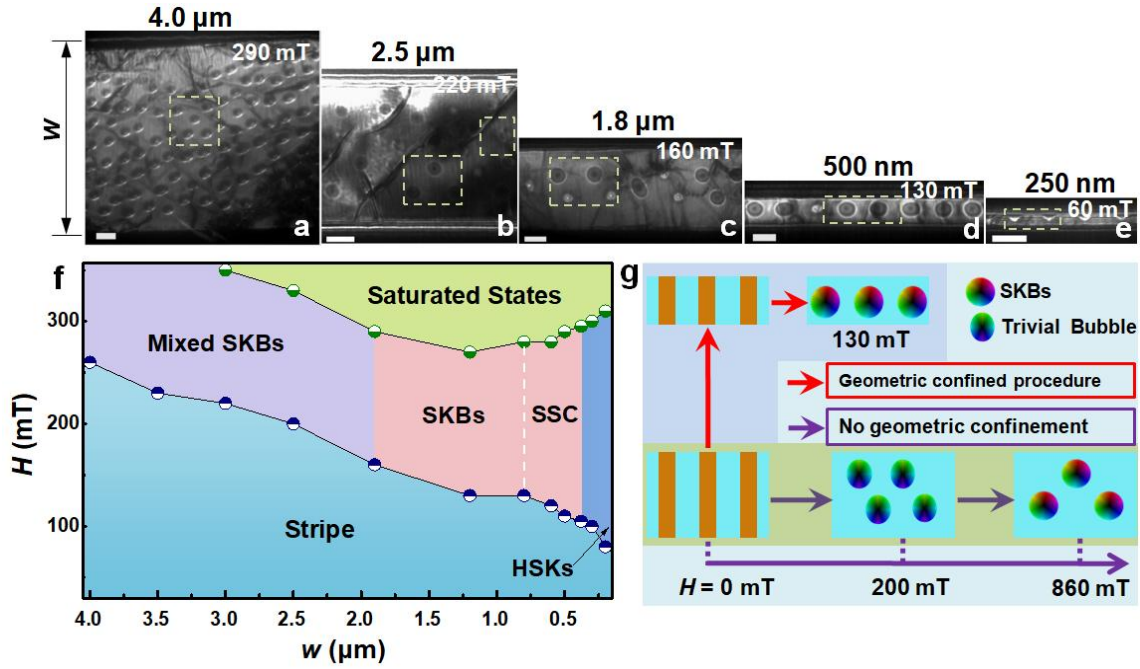

**Figure 3. Experimental spin structures phase diagram for $Fe_3Sn_2$ nanostripes at 300K. (a-e)** The under-focused LTEM images for $Fe_3Sn_2$ nanostripes with various widths under their corresponding critical field $H_c$, where the stripe domains completely transform into trivial bubbles or skyrmion bubbles. The black lines with arrows show the width of the nanostripe. The regions in the white boxes present the different types of magnetic bubble domains. The scale bar in **a-e** is 500nm. **(f)** Experimental spin structure phase diagram for $Fe_3Sn_2$ nanostripes as a function of magnetic fields and widths. The blue-white dots represent the lower critical field and the green-white dots represent the upper critical field $H_u$ where the bubbles or skyrmion bubbles just transformed into saturated states completely. The colored squares correspond to different magnetic domains. Depending on the width, stripe domains can evolve into four different magnetic states: mixed SKBs (coexistence of trivial bubbles, metastable SKBs, and SKBs), SKBs, SSC (single SKB chain), and half-skyrmion-like states (HSKs). The white dash line suggests the boundary where a multiple skyrmion chain transforms into a single skyrmion chain (SSC). **(g)** Schematic diagram illustrating the effects of geometric confinement on the evolution of magnetic domains in $Fe_3Sn_2$. Two different experimental paths are mapped out with two series of colored arrows. The purple channel shows the conventional transformation process of domains under different out-of-plane magnetic fields in the sample, without geometric confinement. The red channel presents the new evolutionary procedure of magnetic domains, observed in the width-confined $Fe_3Sn_2$ nanostripes.

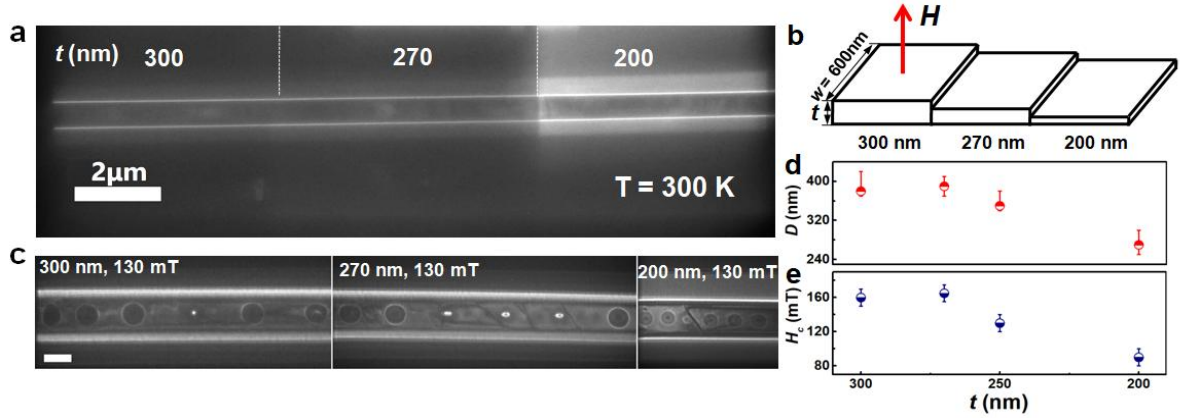

**Figure 4. Thickness-dependent spin structures and sizes of SKBs.** (**a**) TEM image of a $w \sim 600$ nm nanostripe with different thicknesses. Thickness differences are represented as different levels of contrast (separated by white dash lines). (**b**) Schematic illustration of the stage-shaped nanostripe. $H$ represents the applied field. (**c**) under-focused LTEM images of the regions with different thicknesses under an external of magnetic field of 130 mT at 300 K. Scale bar is 500 nm. (**d**) Thickness $t$ dependence of the diameter $D$ of a SKB under a magnetic field of 140 mT and the critical field $H_c$ for the $w \sim 600$ nm nanostripe at 300 K. The values of $D$ for the thickness of 300 nm, 270 nm, and 200 nm are obtained from the $w \sim 600$ nm stage-shaped sample, while $D$ for the thickness of 250 nm was deduced from the nanostripe shown in Fig.1. The error bars were added based on the size of SKBs observed in the stage sample and the nanostripe shown in Fig. 2.

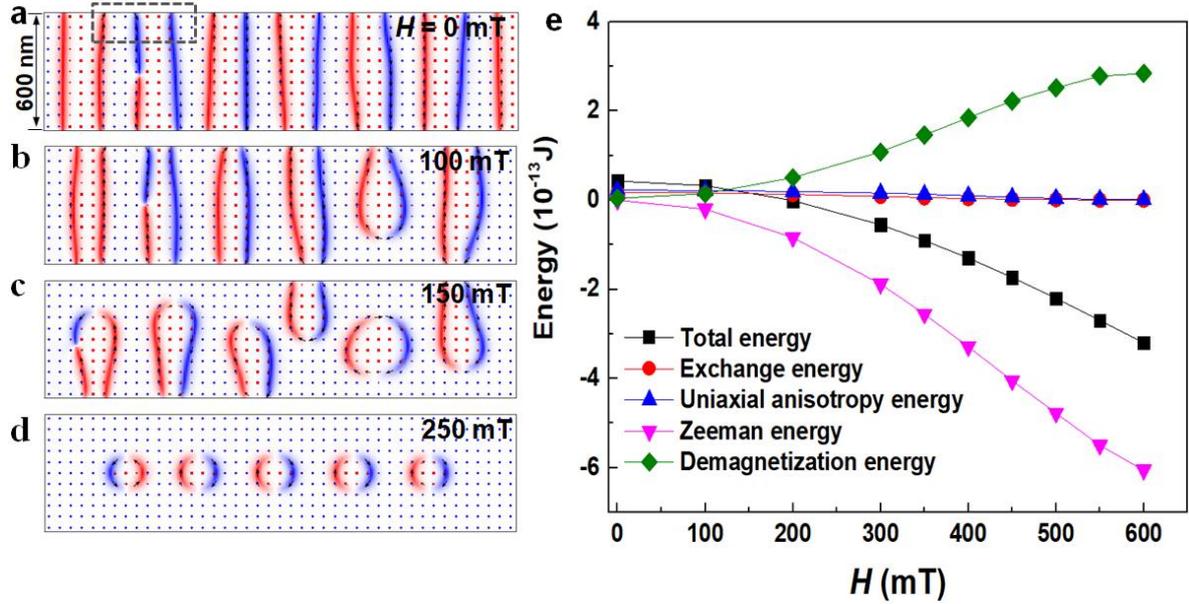

**Figure 5. Simulated magnetization process for a $w \sim 600$ nm, $t \sim 250$ nm nanostripe. (a-d)** The micromagnetic simulations of domain magnetization process as a function of the magnetic field for the $w \sim 600$ nm, $t \sim 250$ nm nanostripe. The in-plane magnetization along the *y*-axis ($m_y$) is represented by pixels in red (+$m_y$) and blue (-$m_y$), whereas the *z*-axis magnetization is represented by the white pixels. The region enclosed by the black box in (a) demonstrates the spins helixes with opposite directions of magnetization slightly attract each other while the ones with the same direction repel each other. **(e)** The field dependence of the energy terms in the simulations for the width $w \sim 600$ nm, and thickness $t \sim 250$ nm nanostripe. With the increase of magnetic field, the uniaxial anisotropy energy and exchange energy change nearly unchanged, whereas the Zeeman energy and demagnetization energy increases significantly. Notably, in the simulations, the demagnetization energy represents the energy term of dipole-dipole interaction.

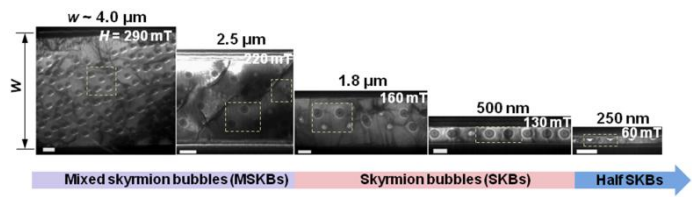

For Table of Contents Only